\documentclass{PoS}%
\usepackage{amsmath}
\usepackage{amsfonts}
\usepackage{amssymb}
\usepackage{graphicx}%
\providecommand{\U}[1]{\protect\rule{.1in}{.1in}}

\title{%
{\protect \vspace{-5mm}
\normalsize \begin{flushright}
\begin{minipage}{4cm}
 KEK Preprint 2016-59 \\
CHIBA-EP-221
\end{minipage}
\end{flushright}
\vspace{5mm} }
Quark confinement to be caused by Abelian or non-Abelian dual superconductivity in the SU(3) Yang-Mills theory
}

\ShortTitle{Quark confinement to be caused by Abelian or non-Abelian dual superconductivity}

\author{\speaker{Akihiro Shibata}\\
        Computing Research Center, High Energy Accelerator Research Organization (KEK) \\
        E-mail: \email{Akihiro.Shibata@kek.jp}}

\author{Kei-Ichi Kondo\\
Department of Physics, Graduate School of Science, Chiba University, Chiba 263-8522, Japan \\
E-mail: \email{kondok@faculty.chiba-u.jp}}

\author{Seikou Kato\\
Oyama National College of Technology, Oyama, Tochigi 323-0806, Japan \\
E-mail:\email{skato@oyama-ct.ac.jp}}

\author{Toru Shinohara\\
Department of Physics, Graduate School of Science, Chiba University, Chiba 263-8522, Japan \\
E-mail:\email{sinohara@graduate.chiba-u.jp}}

\abstract{
The dual superconductivity is a promising mechanism for quark confinement. 
We have presented a new formulation of the Yang-Mills theory on the lattice 
that enables us to change the original non-Abelian gauge field into the new field variables 
such that one of them called the restricted field gives the dominant contribution 
to quark confinement in the gauge independent way. 
We have pointed out that the SU(3) Yang-Mills theory has another reformulation 
using new field variables (minimal option), 
in addition to the way adopted by Cho, Faddeev and Niemi (maximal option).
In the past lattice conferences, we have shown the numerical evidences 
that support the non-Abelian dual superconductivity using the minimal 
option for the SU(3) Yang-Mills theory. This result should be compared 
with Abelian dual superconductivity obtained in the maximal option which is  
a gauge invariant extension of the conventional Abelian projection method in the maximal Abelian gauge.

In this talk, we focus on discriminating between two reformulations, i.e., maximal and minimal options of 
the $SU(3)$ Yang-Mills theory from the viewpoint of dual superconductivity for quark confinement. 
We investigate the confinement/deconfinement phase transitions at finite temperature in both options, 
which are compared with the original Yang-Mills theory. 
For this purpose, we measure the distribution of Polyakov-loops and the Polyakov-loop average,
the correlation function of the Polyakov loops and the distribution of the chromoelectric flux connecting a quark 
and antiquark in both confinement and deconfinement phases.
}

\FullConference{34th annual International Symposium on Lattice Field Theory\\
		24-30 July 2016\\
		University of Southampton, UK}

\begin{document}
\section{Introduction}

The dual superconductivity is a promising mechanism for quark confinement
\cite{DualMeisser75}. In order to establish this picture, we have presented a
new formulation of the Yang-Mills theory on the lattice that enables us to
decompose the original Yang-Mills (YM) gauge link valiable $U_{x,\mu}$ into
the gauge link variable\ $V_{x,\mu}$ corresponding to its maximal stability
subgroup of the gauge group and the remainder $X_{x,\mu}$, $U_{x,\mu}%
=X_{x,\mu}V_{x,\mu}$, where the restricted field $V_{x,\mu}$ could be the
dominant mode for quark confinement. (For a review see \cite{PhysRep}). For
the $SU(3)$ YM theory, we have two options of the decomposition: the minimal
and maximal options. In the minimal option, especially, the maximal stability
group is non-Abelian $U(2)\,$and the restricted field contains the non-Abelian
magnetic monopole. In the preceding works, we have shown numerical evidences
of the non-Abelian dual superconductivity using the minimal option for the
$SU(3)$ YM theory on a lattice: the restricted field and the extracted
non-Abelian magnetic monopole dominantly  reproduces the string tension in the
linear potential of the $SU(3)$ YM theory \cite{abeliandomSU(3)}, and the
$SU(3)$ YM vacuum is the type I dual superconductor detected by the
chromoelectric flux tube and the magnetic monopole current induced around it,
which is a novel feature obtained by our simulations \cite{DMeisner-TypeI2013}%
. We have further investigated the \ confinement/deconfinement phase
transition in view of this non-Abelian dual superconductivity
picture\cite{lattice2013}\cite{lattice2014}\cite{SCGT15}\cite{lattice2015}.
These results should be compared with the maximal option which is adopted
first by Cho, Faddeev, and Niemi \cite{CFNS-C}. The maximal stablity group is
Abelian $U(1)\times U(1)$ and the restricted field involves only the Abelian
magnetic monopole \cite{lattce2007}\cite{ChoKundy2014}. This is nothing but
the gauge invariant extension of the Abelian projection in the maximal Abelian
gauge\cite{Suganuma}\cite{suganuma-sakumichi}.

In this talk, we focus on discriminating between two reformulations, i.e.,
maximal and minimal options of the $SU(3)$ YM theory  from the viewpoint of
dual superconductivity for quark confinement. For this purpose, we measure
string tension for the restricted non-Abelian field of both minimal and
maximal option in comparison with the string tension for the original YM
field. We also investigate the dual Meissner effect by measuring the
distribution of the chromoelectric flux connecting a quark and an antiquark
and the induced magnetic-monopole current around the flux tube. 

\section{Gauge link decompositions}

Let $U_{x,\mu}=X_{x,\mu}V_{x,\mu}$ be a decomposition of the YM link variable
$U_{x,\mu}$, where the YM field and the decomposed new variables are
transformed by full $SU(3)$ gauge transformation $\Omega_{x}$ such that
$V_{x,\mu}$ is transformed as a gauge link variable and $X_{x,\mu}$ as a site
variable \cite{exactdecomp}:
\begin{subequations}
\label{eq:gaugeTransf}%
\begin{align}
U_{x,\mu} &  \longrightarrow U_{x,\mu}^{\prime}=\Omega_{x}U_{x,\mu}%
\Omega_{x+\mu}^{\dag},\\
V_{x,\mu} &  \longrightarrow V_{x,\mu}^{\prime}=\Omega_{x}V_{x,\mu}%
\Omega_{x+\mu}^{\dag},\text{ \ }X_{x,\mu}\longrightarrow X_{x,\nu}^{\prime
}=\Omega_{x}X_{x,\mu}\Omega_{x}^{\dag}.
\end{align}
For the SU(3) YM theory, we have two options discriminated by its stability
group, so called the minimal and maximal options.

\subsection{Minimal option}

The minimal option is obtained for the stability subgauge group $\tilde
{H}=U(2)=SU(2)\times U(1)\subset SU(3)$. By introducing the color field
$\mathbf{h}_{x}=\xi(\lambda^{8}/2)\xi^{\dag}$ $\in SU(3)/U(2)$ with
$\lambda^{8}$ being the last diagonal Gell-Mann matrix and $\xi$ an $SU(3)$
group element, the decomposition is given by the defining equation:
\end{subequations}
\begin{subequations}
\label{eq:def-min}%
\begin{align}
&  D_{\mu}^{\epsilon}[V]\mathbf{h}_{x}:=\frac{1}{\epsilon}\left[  V_{x,\mu
}\mathbf{h}_{x+\mu}-\mathbf{h}_{x}V_{x,\mu}\right]  =0,\label{eq:def1-min}\\
&  g_{x}:=e^{i2\pi q/3}\exp(-ia_{x}^{0}\mathbf{h}_{x}-i\sum\nolimits_{j=1}%
^{3}a_{x}^{(j)}\mathbf{u}_{x}^{(j)}).\label{eq:def2-min}%
\end{align}
Here, the variable $g_{x}$ is an undetermined parameter from
Eq.(\ref{eq:def1-min}), $\mathbf{u}_{x}^{(j)}$ 's are $su(2)$-Lie algebra
valued, and \thinspace$q$ has an integer value. These defining equations can
be solved exactly, and the solution is given by
\end{subequations}
\begin{subequations}
\label{eq:decomp-min}%
\begin{align}
X_{x,\mu} &  =\widehat{L}_{x,\mu}^{\dag}\det(\widehat{L}_{x,\mu})^{1/3}%
g_{x}^{-1},\text{ \ \ \ }V_{x,\mu}=X_{x,\mu}^{\dag}U_{x,\mu},\\
\widehat{L}_{x,\mu} &  :=\left(  L_{x,\mu}L_{x,\mu}^{\dag}\right)
^{-1/2}L_{x,\mu},\\
\text{\ }L_{x,\mu} &  :=\frac{5}{3}\mathbf{1}+\frac{2}{\sqrt{3}}%
(\mathbf{h}_{x}+U_{x,\mu}\mathbf{h}_{x+\mu}U_{x,\mu}^{\dag})+8\mathbf{h}%
_{x}U_{x,\mu}\mathbf{h}_{x+\mu}U_{x,\mu}^{\dag}\text{ .}%
\end{align}
Note that the above defining equations correspond to the continuum version:
$D_{\mu}[\mathcal{V}]\mathbf{h}(x)=0$ and $\mathrm{tr}(\mathbf{h}%
(x)\mathcal{X}_{\mu}(x))$ $=0$, respectively. In the naive continuum limit, we
have reproduced the decomposition $\mathbf{A}_{\mathbf{\mu}}(x)=\mathbf{V}%
_{\mu}(x)+\mathbf{X}_{\mu}(x)$ in the continuum theory \cite{KSM05}.

The decomposition is uniquely obtained as the solution (\ref{eq:decomp-min})
of Eqs.(\ref{eq:def-min}), if color fields$\{\mathbf{h}_{x}\}$ are obtained.
To determine the configuration of color fields, we use the reduction condition
to formulate the new theory written by new variables ($X_{x,\mu}$,$V_{x,\mu}$)
which is equipollent to the original YM theory. Here, we use the reduction
functional:
\end{subequations}
\begin{equation}
F_{\text{red}}[\mathbf{h}_{x}]=\sum\nolimits_{x,\mu}\mathrm{tr}\left\{
(D_{\mu}^{\epsilon}[U_{x,\mu}]\mathbf{h}_{x})^{\dag}(D_{\mu}^{\epsilon
}[U_{x,\mu}]\mathbf{h}_{x})\right\}  , \label{eq:reduction-min}%
\end{equation}
and then color fields $\left\{  \mathbf{h}_{x}\right\}  $ are obtained by
minimizing the functional (\ref{eq:reduction-min}).

\subsection{Maximal option}

The maximal option is obtained for the stability subgroup of the maximal tarus
group $\tilde{H}=U(1)\times U(1)\subset SU(3)$, and the resulting
decomposition is the gauge-invariant extension of the Abelian projection in
the maximal Abelian i'l'`j gauge. By introducing the
color field $\mathbf{n}^{(3)}=\xi(\lambda^{3}/2)\xi^{\dag}$,$\ \mathbf{n}%
^{(8)}=$ $\xi(\lambda^{8}/2)\xi^{\dag}$ $\in SU(3)/U(2)$ with $\lambda
^{3},\lambda^{8}$ being the two diagonal Gell-Mann matrices and $\xi$ an
$SU(3)$ group element,\ the decomposition is given by the defining equation:
\begin{subequations}
\label{eq:Defeq-max}%
\begin{align}
&  D_{\mu}^{\epsilon}[V]\mathbf{n}_{x}^{(j)}:=\frac{1}{\epsilon}\left[
V_{x,\mu}\mathbf{n}_{x+\mu}^{(j)}-\mathbf{n}_{x}^{(j)}V_{x,\mu}\right]
=0\text{ \ \ \ }j=3,8,\label{eq:def-max-1}\\
&  g_{x}:=e^{i2\pi q/3}\exp(-ia_{x}^{3}\mathbf{n}_{x}^{(3)}-ia_{x}%
^{(8)}\mathbf{n}_{x}^{(8)}).
\end{align}
The variable $g_{x}$ is an undetermined parameter from Eq.(\ref{eq:def-max-1}%
), and $q$ has an integer value. These defining equations can be solved
exactly, and the solution is given by
\end{subequations}
\begin{subequations}
\label{eq:decomp_max}%
\begin{align}
X_{x,\mu} &  =\widehat{K}_{x,\mu}^{\dag}\det(\widehat{K}_{x,\mu})^{1/3}%
g_{x}^{-1},\text{ \ \ \ }V_{x,\mu}=X_{x,\mu}^{\dag}U_{x,\mu},\\
\widehat{K}_{x,\mu} &  :=\left(  K_{x,\mu}K_{x,\mu}^{\dag}\right)
^{-1/2}K_{x,\mu},\\
K_{x,\mu} &  :=\mathbf{1}+6(\mathbf{n}_{x}^{(3)}U_{x,\mu}\mathbf{n}_{x+\mu
}^{(3)}U_{x,\mu}^{\dag})+6(\mathbf{n}_{x}^{(8)}U_{x,\mu}\mathbf{n}_{x+\mu
}^{(8)}U_{x,\mu}^{\dag})
\end{align}
Note that the above defining equations correspond to the continuum version:
$D_{\mu}[\mathcal{V}]\mathbf{n}^{(j)}(x)=0$ and $\mathrm{tr}(\mathbf{n}%
^{(j)}(x)\mathcal{X}_{\mu}(x))$ $=0$, respectively. In the naive continuum
limit, we have reproduced the decomposition $\mathbf{A}_{\mathbf{\mu}%
}(x)=\mathbf{V}_{\mu}(x)+\mathbf{X}_{\mu}(x)$ in the continuum theory.( See
\cite{KSM05}\cite{CFNS-C}.)

To determine the configuration of color fields, we use the reduction condition
to formulate the new theory written by new variables ($X_{x,\mu}$,$V_{x,\mu}$)
which is equipollent to the original YM theory. Here, we use the reduction
functional:
\end{subequations}
\begin{equation}
F_{\text{red}}[\mathbf{n}_{x}^{(3)},\mathbf{n}_{x}^{(8)}]=\sum\nolimits_{x,\mu
}\sum\nolimits_{j=3,8}\mathrm{tr}\left\{  (D_{\mu}^{\epsilon}[U_{x,\mu
}]\mathbf{n}_{x}^{(j)})^{\dag}(D_{\mu}^{\epsilon}[U_{x,\mu}]\mathbf{n}%
_{x}^{(j)})\right\}  ,\label{eq:reduction-max}%
\end{equation}
and then color fields $\left\{  \mathbf{n}_{x}^{(3)},\mathbf{n}_{x}%
^{(8)}\right\}  $ are obtained by minimizing the functional
(\ref{eq:reduction-max}). It should be noticed that the maximal option gives
the gauge invariant extension of the Abelian projection in the maximal Abelian
gauge. 

\section{Lattice Data}

We generate the YM gauge field configurations (link variables) $\{U_{x,\mu}\}$
using the standard Wilson action on the lattice with the size of $L^{3}\times
N_{T}=24^{3}\times6$. We prepare 500 data sets  at $\beta:=2N_{c}%
/g^{2}\ (N_{c}=3)=5.8,5.9,6.0,6.1,6.2,6.3$ every 500 sweeps after 10000
thermalization. The temperature is controlled by changing the parameter
$\beta$. We obtain two types of decomposed gauge link variables $U_{x,\mu
}=X_{x,\mu}V_{x,\mu}$ for each gauge link by using the formula
Eqs.(\ref{eq:decomp-min}) and (\ref{eq:decomp_max}) given in the previous
section, after the color-field configuration $\{\mathbf{h}_{x}\}$ and
$\left\{  \mathbf{n}_{x}^{(3)},\mathbf{n}_{x}^{(8)}\right\}  $ are obtained by
solving the reduction condition as minimizing  the functional
(\ref{eq:reduction-min}) and (\ref{eq:reduction-max}), respectively. In the
measurement of the Wilson loop average defined below, we apply the APE
smearing technique to reduce noises.%

\begin{figure}[tbp] \centering
\begin{minipage}{\textwidth} \centering
\includegraphics[width=4.9cm]{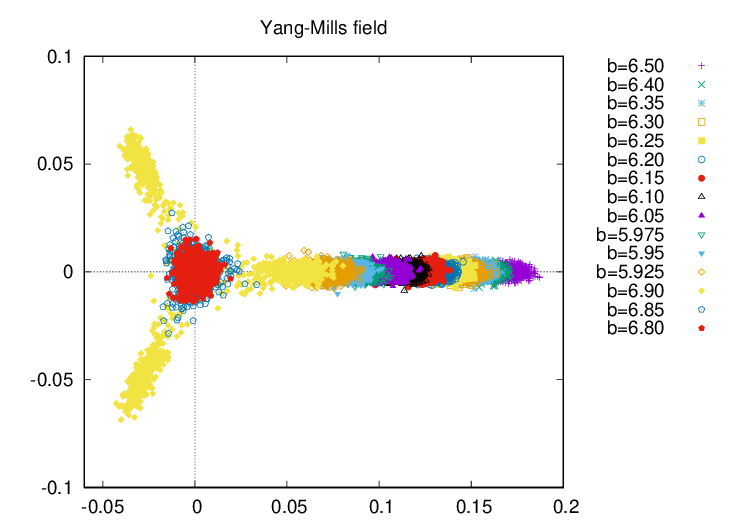}
\includegraphics[width=4.8cm]{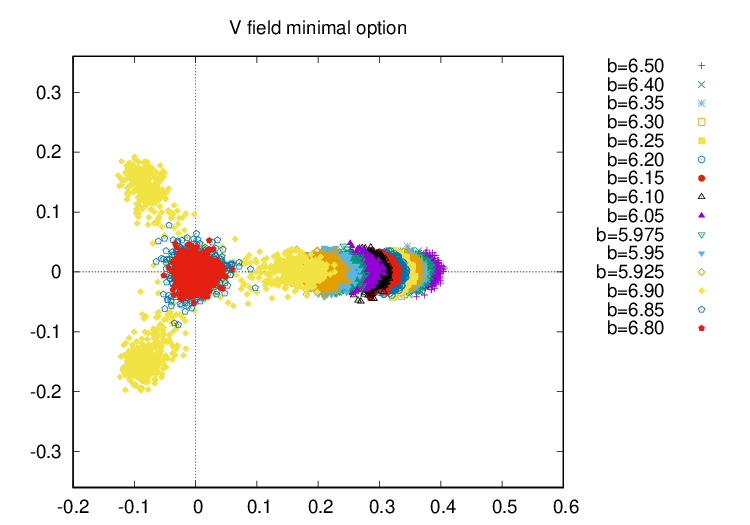}
\includegraphics[width=4.8cm]{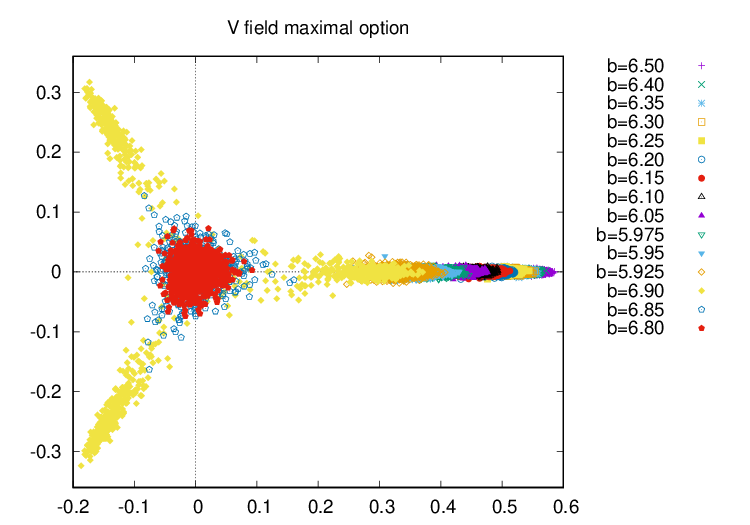}%
\caption{
Distribution of the space-averaged Polyakov loops:
The space-averaged Polyakov loop eq({\protect \ref{eq:Ploop-Space-ave}})
 for each configuration are  plotted in the complex plane.
(a) original (YM) field, (b) restricted field in the minimal option, (c) restricted field in the maximal option.
}%
\label{fig:ploop}%
\end{minipage}%
\end{figure}
\begin{figure}[tbp] \centering
\begin{minipage}{\textwidth} \centering
\includegraphics[width=7.0cm]{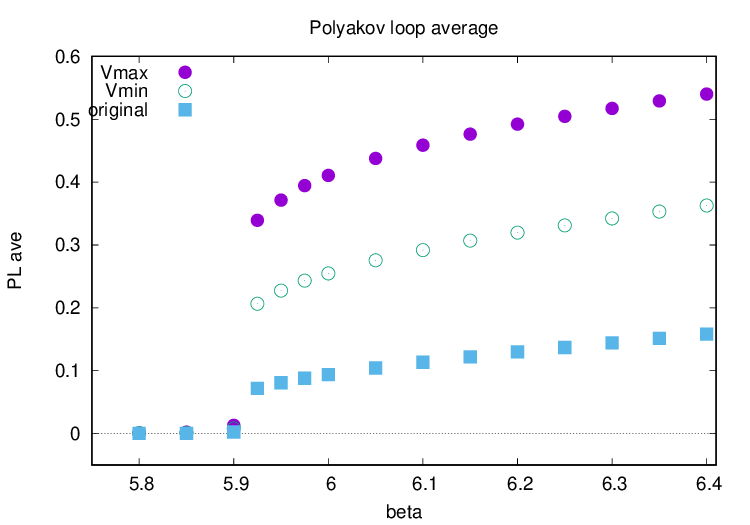}
\includegraphics[width=7.0cm]{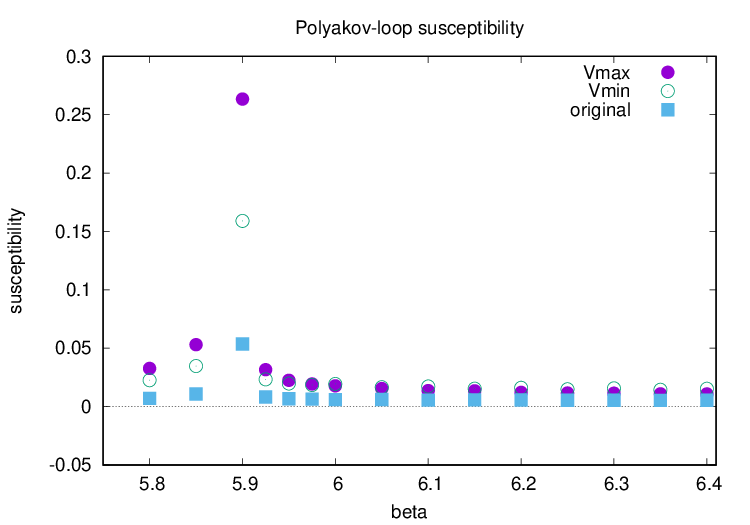}\caption{
(Left) The combination plot of the Polyakov-loop averages for the original (YM) field,  minimal option maximal option
from bottom to top.
(Right)  The combination plot of the suseceptability of  the Polyakov-loop for the original field, minimal option
and maximal option.
}\label{fig:ploop_ave}%
\end{minipage}%
\end{figure}%

First, we investigate the distribution of a single Polyakov loop $P_{\ast
}(\mathbf{x)}$ in both options as well as the original Yang-Mills theory:%
\begin{equation}
P_{YM}(\mathbf{x}):=\text{\textrm{tr}}\left(  P\prod\limits_{t=1}^{N_{T}%
}U_{(\mathbf{x},t),4}\right)  \text{, \ }P_{\min}(\mathbf{x}%
):=\text{\textrm{tr}}\left(  P\prod\limits_{t=1}^{N_{T}}V_{(\mathbf{x}%
,t),4}^{(\min)}\text{ }\right)  \text{\ \ ,}P_{\max}(\mathbf{x}):=\sum
_{\mathbf{x}}\text{\textrm{tr}}\left(  P\prod\limits_{t=1}^{N_{T}%
}V_{(\mathbf{x},t),4}^{(\max)}\right)  ,\text{ }\label{eq;Polyakov-loop}%
\end{equation}
Figure \ref{fig:ploop} represents the distribution of the space-averaged
Polyakov loop $P_{\ast}(\mathbf{x)}$ defined by
\begin{equation}
P_{YM}=\frac{1}{L^{3}}\sum_{\mathbf{x}}P_{YM}(\mathbf{x}),\ \ \ P_{\min}%
=\frac{1}{L^{3}}\sum_{\mathbf{x}}P_{\min}(\mathbf{x}),\text{ \ \ }P_{\max
}=\frac{1}{L^{3}}\sum_{\mathbf{x}}P_{\max}(\mathbf{x}%
),\label{eq:Ploop-Space-ave}%
\end{equation}
for each configuration in the complex plane. The value of the space-averaged
Polyakov loop are different among the options, but all the distributions
equally refrect the expected center symmetry $Z(3)$ of $SU(3)$. The left panel
of Figure \ref{fig:ploop_ave} shows the Polyakov-loop average, $\left\langle
P_{YM}\right\rangle $, $\left\langle P_{\min}\right\rangle $, $\left\langle
P_{\max}\right\rangle $.  The Polyakov-loop average is an order parameter of
the center symmetry breaking and restoration, which is the conventional order
parameter of the confinement/deconfinement phase transition. The right panel
of Fig.\ref{fig:ploop_ave}$\,\ $shows the susceptibility.
Fig.\ref{fig:ploop_ave} shows that three Polyakov-loop average give the same
critical point $\beta=5.9$. Therefore, both the minimal and maximal options
reproduce the critical point in the original Yang-Mills field.%

\begin{figure}[tbp] \centering
\begin{minipage}{\textwidth} \centering
\includegraphics[width=15.5cm]{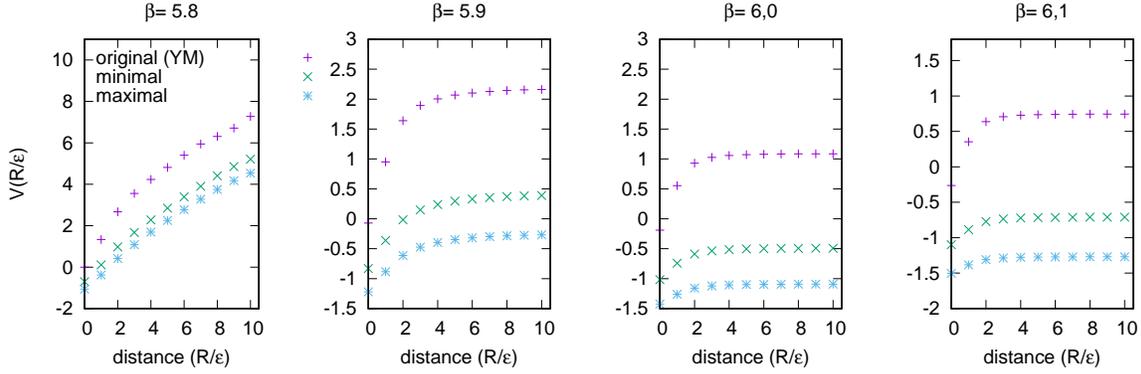}
\caption{@The two-point correlation functions of the Polyakov loops
calculated from the  Yang-Mills field and V-fields in the minimal
and maximal options are ploted in the same panel for various values of $\beta$
} \label{fig:potential2}%
\end{minipage}%
\end{figure}

Next, we investigate the two-point correlation function of the Polyakov loops,
which is related to the mixed static-potential of the singlet and adjoint
composite states in the following way \cite{Yangi-Hatsuda05}:
\begin{equation}
V_{U}(r=|\mathbf{x}-\mathbf{y}|)=\log(\left\langle P_{U}(\mathbf{x}%
)P_{U}(\mathbf{y})\right\rangle )\text{, \ \ \ \ }\left\langle P_{U}%
(\mathbf{x})P_{U}(\mathbf{y})\right\rangle \simeq\frac{1}{N_{c}}%
e^{-F(S)/T}+\frac{N_{c}^{2}-1}{N_{c}}e^{-F(A)/T}.\label{eq:Potential}%
\end{equation}
Figure \ref{fig:potential2} shows the combination plot eq.(\ref{eq:Potential})
for various temperature ($\beta$). In the both options we find that the
restricted field ($V$-field) is dominant for the Polyakov loop correlation
function  and reproduces the static potential of the original YM theory in the
long distance $r=|\mathbf{x}-\mathbf{y}|$.

\begin{figure}[tbp] \centering
\begin{minipage}{\textwidth} \centering
\vspace{-5mm} \includegraphics[height=46mm,angle=270]{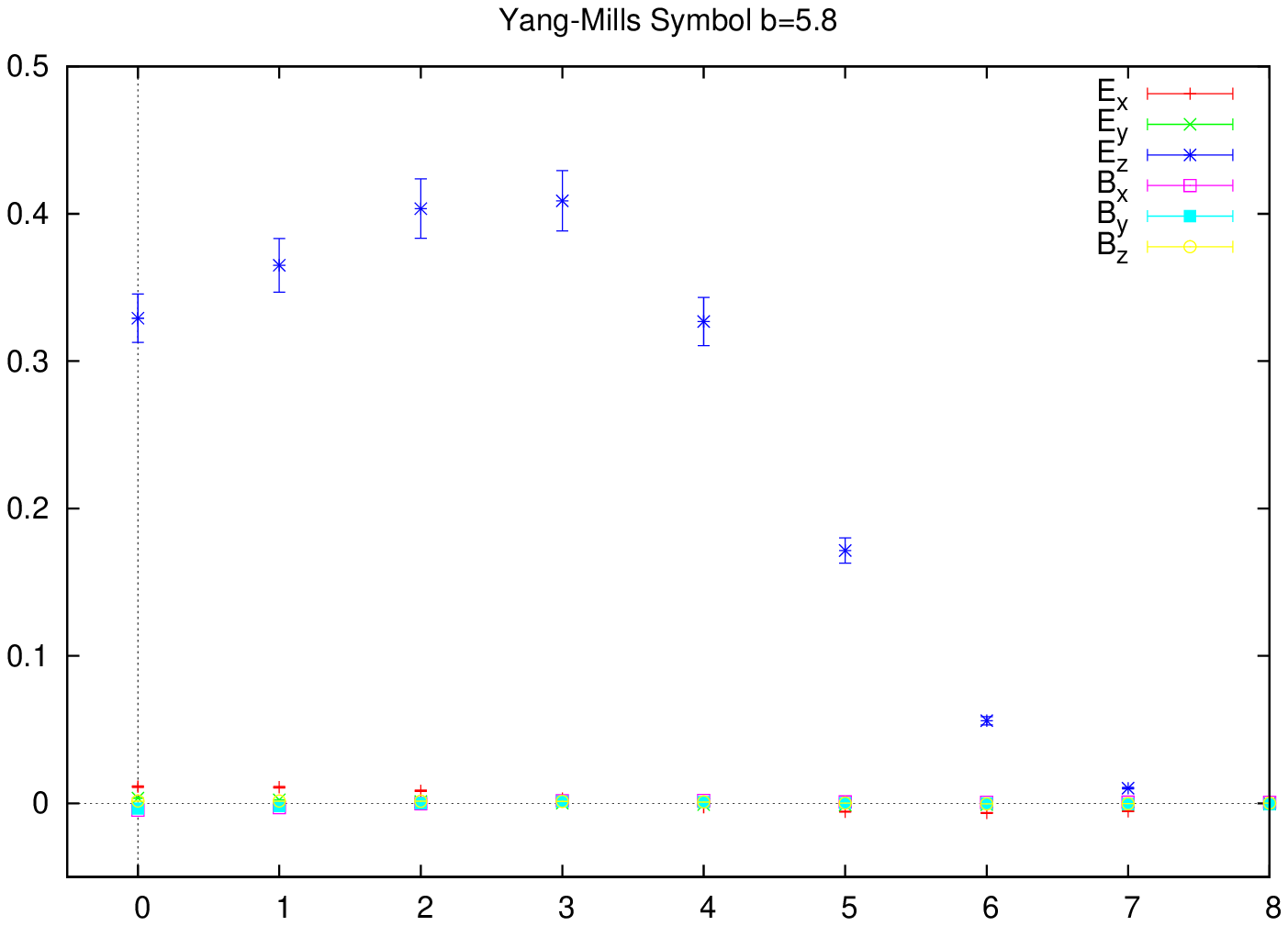}
\includegraphics[height=46mm,angle=270]{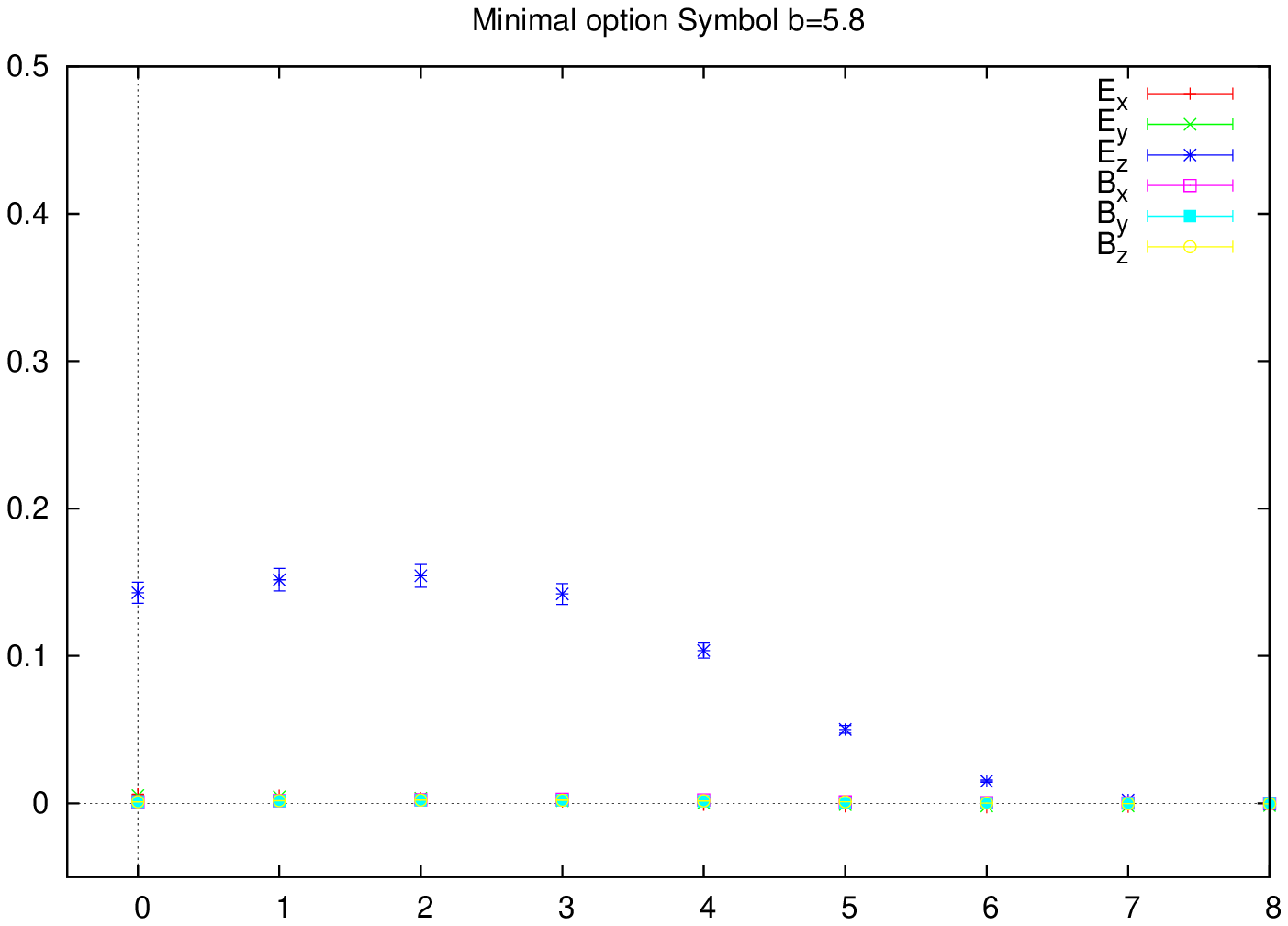}
\includegraphics[height=46mm,angle=270]{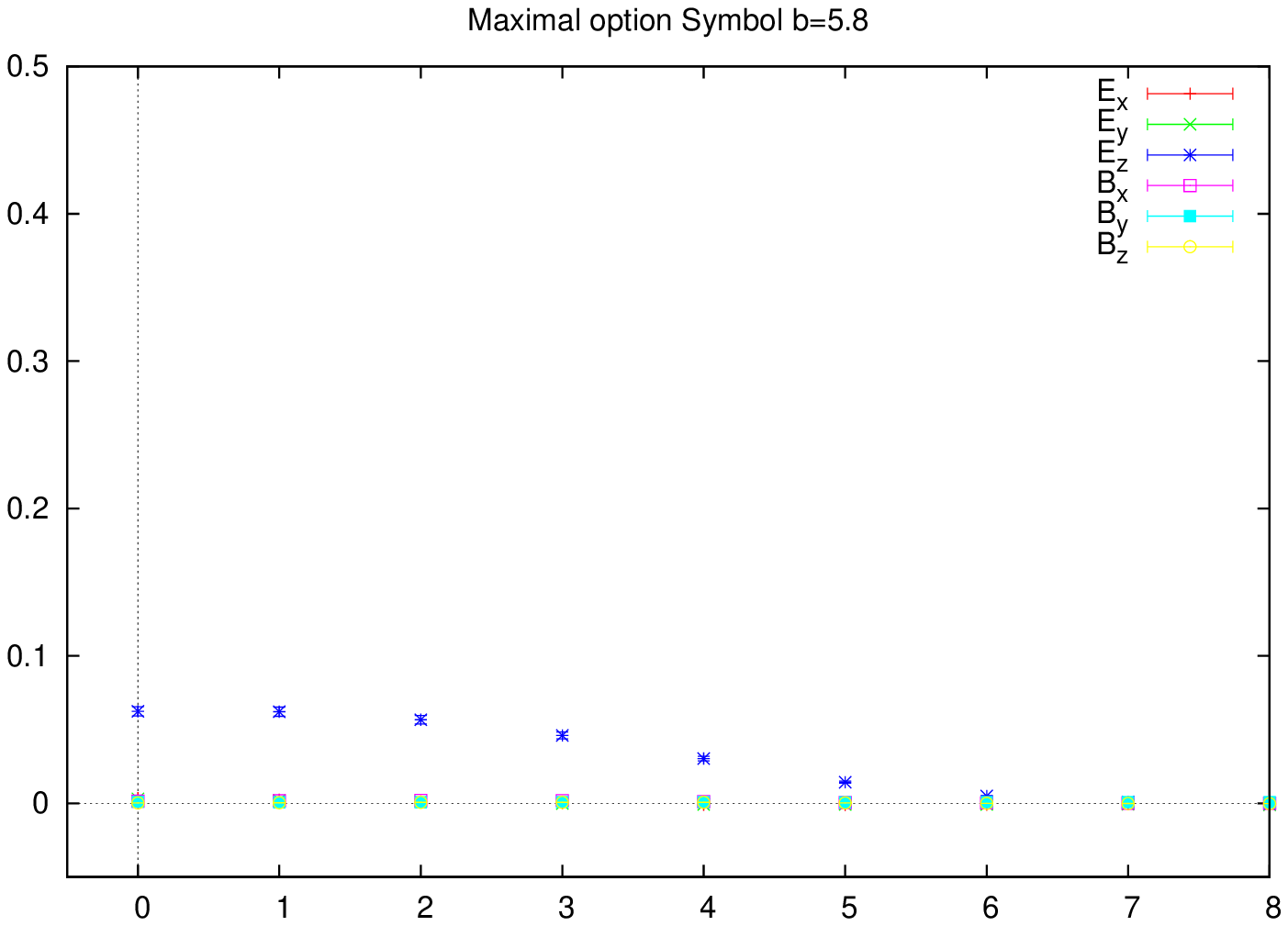}
\includegraphics[height=46mm,angle=270]{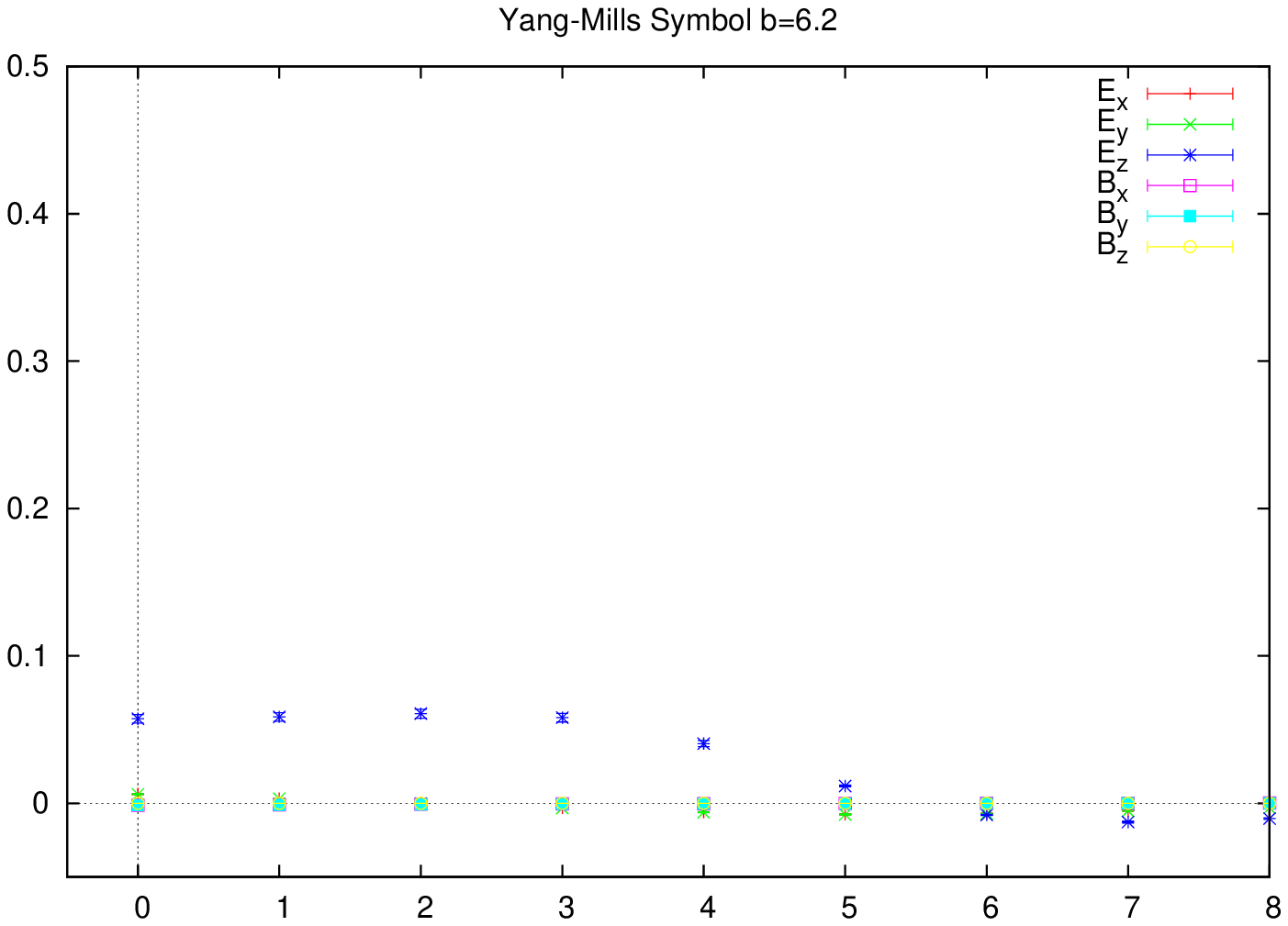}
\includegraphics[height=46mm,angle=270]{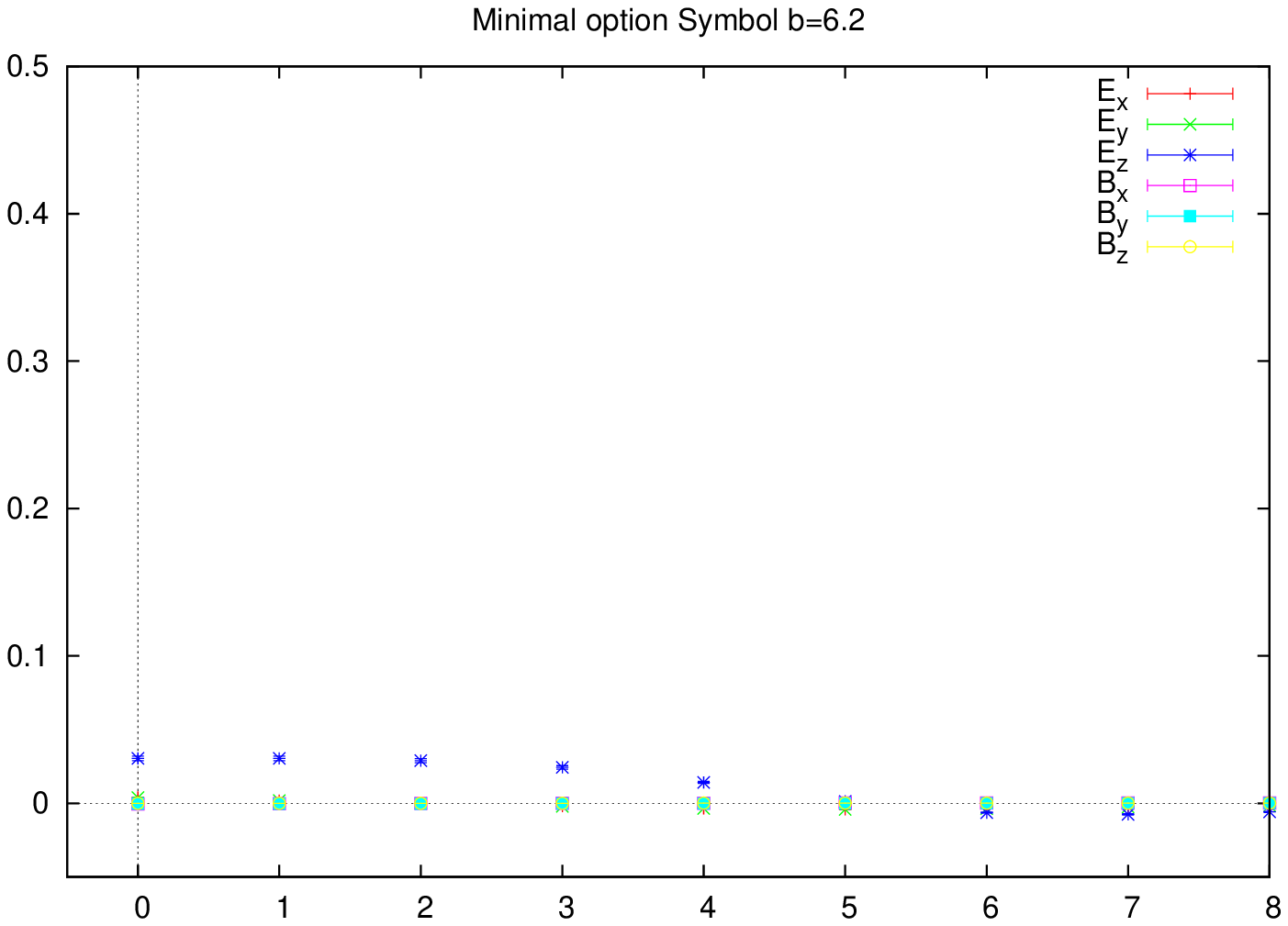}
\includegraphics[height=46mm,angle=270]{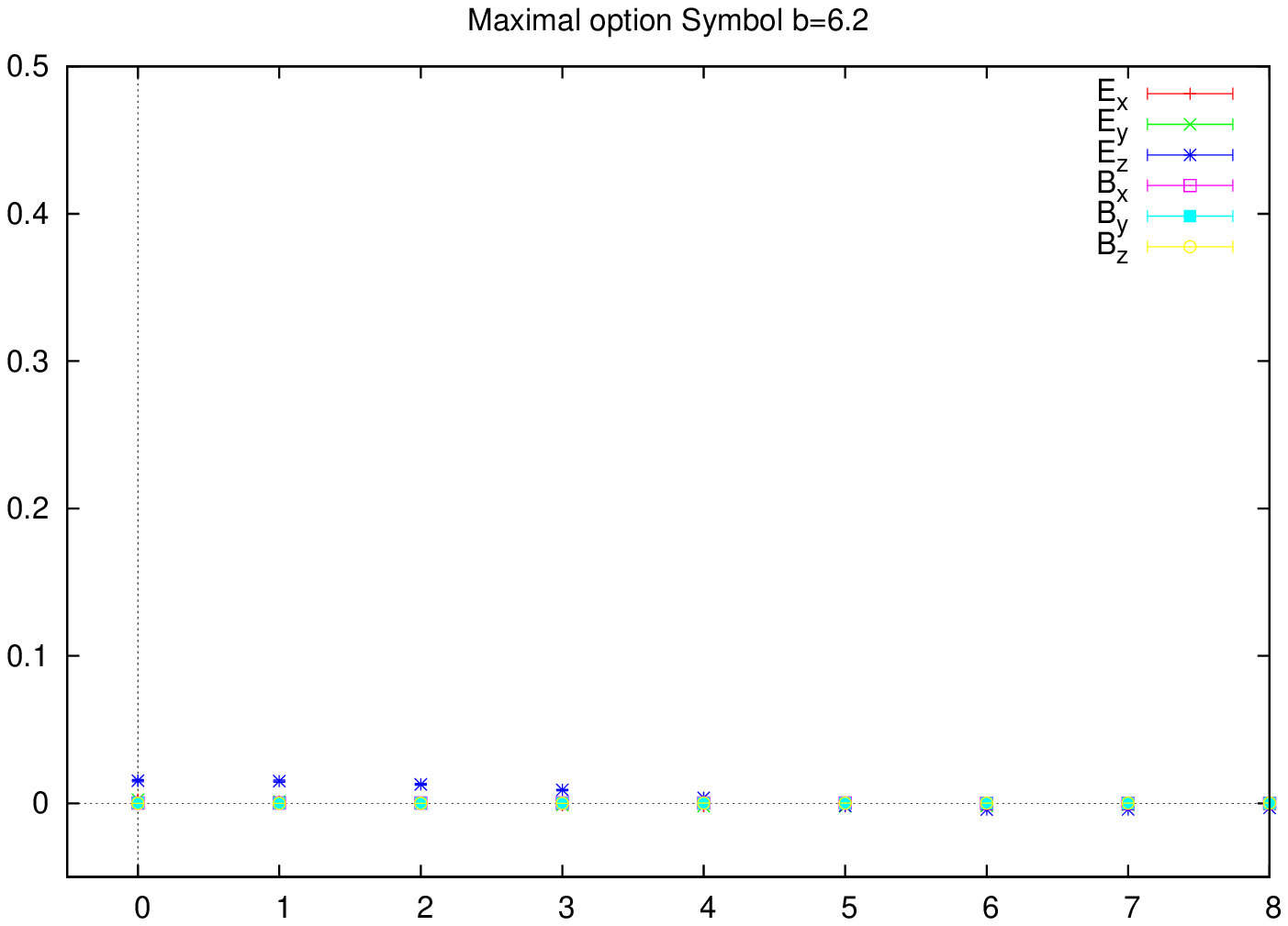}
\caption{ Chromo flux created by a pair of quark and anti-quark.
The flux is measured at the point of distance y from the 1/3 dividing  point of
quark and antiquark lying on the z-axis.
Upper and lower panels represent chromo flux in confining phase ($\beta=5.8$)
and deconfining phase ($\beta=6.2$), respectively.
Each panel in the left midle and right represents the measurement of the original (YM) field,
the minimal and the maximal options, respectively.
} \label{fig:flux-confine}%
\end{minipage}%
\end{figure}

Finally, we study the dual Meissner effect. For this purpose, we measure the
chromo flux at finite temperature created by a quark-antiquark pair which is
represented by the maximally extended Wilson loop $W$ of the $R\times T$
rectangle, i.e., the field strength of the chromo flux at the position $P$ is
measured by using a plaquette variable $U_{p}$ at $P$ as the probe operator
for the field strength \cite{Giacomo}:
\begin{equation}
\rho_{_{U_{P}}}:=\frac{\left\langle \mathrm{tr}\left(  WLU_{p}L^{\dag}\right)
\right\rangle }{\left\langle \mathrm{tr}\left(  W\right)  \right\rangle
}-\frac{1}{3}\frac{\left\langle \mathrm{tr}\left(  U_{p}\right)
\mathrm{tr}\left(  W\right)  \right\rangle }{\left\langle \mathrm{tr}\left(
W\right)  \right\rangle },\label{eq:Op}%
\end{equation}
where $L$ is the Schwinger line connecting the source $W$ and the probe
$U_{p}$ needed to obtain the gauge-invariant result (See \cite{lattice2014}).
To measure the chromo flux for the restricted fields of the minimal and
maximal options, we use the operator that the Schwinger line $L$ and the probe
$U_{p}$ are made of the restricted fields. Figure \ref{fig:flux-confine} shows
the preliminary measurements of the chromo flux the both options and the
original field. We find that the chromo-flux tube appears in the confining
phase ($\beta=5.8$), while in the deconfining phase ($\beta=6.2$) the flux
tube disappears, that is to say, the confinement/deconfinement phase
transition is understood as the dual Meissner effect. This is the case for
both options and the original YM field, although the precise profiles of the
chromo flux are different option by option.%

\section{Summary}

By using a new formulation of YM theory, we have investigated possible two
types of the dual superconductivity at finite temperature in the $SU(3)$ YM
theory, i.e., the Non-Abelian dual superconductivity as the minimal option and
the maximal option to be compared with the conventional Abelian dual
superconductivity. In the measurement for both maximal and minimal options as
well as for the original YM\ field at finite temperature, we found the
restricted V-field dominance for both options.\ The restricted fields in the
both options reproduce the center symmetry breaking and restoration of the
original Yang-Mills theory, and give the same critical temperature of the
confining-deconfining phase transition. Then, we have investigated the dual
Meissner effect and found that the chromoelectric flux tube appears in each
option in the confining phase, but it disappears in the deconfining phase.
Thus both options can be adopted as the low-energy effective description of
the original Yang-Mills theory at least within the investigations presented in
this talk.

\subsection*{Acknowledgement}

This work is supported by Grant-in-Aid for Scientific Research (C) 24540252
and 15K05042 from Japan Society for the Promotion Science (JSPS), and in part
by JSPS Grant-in-Aid for Scientific Research (S) 22224003. The numerical
calculations are supported by the Large Scale Simulation Program No.14/15-24
(2014-2015) \ and \ No.15/16-16 (2015-2016) of High Energy Accelerator
Research Organization (KEK).

\end{document}